\documentstyle[12pt,amssymb,cite]{article}

\def\b{\begin{eqnarray}}
\def\e{\end{eqnarray}}
\def\l{\label}

\def\a{\alpha}

\def\n{(\nu,\tilde\nu)}

\begin{document}

\title{KICK ASYMMETRY ALONG A STRONG MAGNETIC FIELD IN THE PROCESS OF
NEUTRINO SCATTERING ON NUCLEONS}
\author{
A.A.\,Gvozdev, I.S.\,Ognev\\[2mm]
\begin{tabular}{l}
Department of Theoretical Physics, Yaroslavl State University,\\
Sovietskaya 14, Yaroslavl 150000, Russia.\\
E-mail: {\em gvozdev@uniyar.ac.ru}, {\em ognev@uniyar.ac.ru}\\ [2mm]
%
%
%
%
\end{tabular}
}

\date{}

\maketitle

ABSTRACT.  The neutrino-nucleon scattering in a collapsing star envelope
with a  strong magnetic field is investigated. The transferred by neutrinos
momentum asymmetry along the field direction is obtained. It is shown that the
neutrino-nucleon scattering gives a contribution to the asymmetry
comparable with direct URCA processes. Hence, the
neutrino-nucleon scattering should be taken into account in estimations
of a possible influence of neutrino reemission processes on a collapsing
star envelope dynamics.  \\[1mm]


{\bf 1. Introduction}\\[1mm]

The most powerful star processes (a Supernova II explosion, a coalescense
of a closed binary system  of neutron stars, an accretion induced
collapse) are of a permanent interest in astrophysics. A collapse in such
systems can lead to the formation of a millisecond remnant
\cite{BK1,W,J1,Sp}. It is usually assumed that the remnant consists of a
compact rigid- rotating core and  a differently  rotating envelope. The
compact core with the typical  size $ R_c \sim 10 km $, the  supranuclear
density $ \rho \gtrsim 10^{13} g / cm^3 $ and the high temperature
$ T \gtrsim 10 MeV $ is opaque to neutrinos.
The remnant envelope with the typical size of several tens of
kilometers, the density $ \rho \sim 10^{11} - 10^{12} g/cm^3 $, and the
temperature $ T \sim 3 - 6 MeV $ is partially transparent to the neutrino
flux. Extremely high neutrino flux with the typical luminosity
$ L_\nu \sim 10^{52} erg/s $ is emitted from the remnant during two or
three seconds after the collapse. As result of a high- rotating
frequency and a medium viscosity of the remnant, a turbulent dynamo and a
large gradient of angular velocities are inevitably produced during a
star contraction.
An extremely strong poloidal magnetic field up to $ B \sim 10^{15} G $
could be generated by a dynamo process in the remnant \cite{Duncan}.
On the other hand, a large gradient of angular velocities in the vicinity
of a rigid- rotating millisecond core can generate a more strong toroidal
magnetic field with the strength $ B \sim 10^{15} - 10^{17} G $ during a
second \cite{BK2}. In the present paper we investigate the influence of
neutrino-nucleon processes on the dynamics of collapsing star millisecond
remnant. \\[2mm]

{\bf 2. Momentum asymmetry in neutrino-nucleon scattering}\\[1mm]

Due to the parity-violation in neutrino-nucleon processes, an asymmetry
along a field direction of macroscopic momentum transferred by neutrinos
to the medium  could exist in an external magnetic field. A quantitative
estimation of the momentum asymmetry is given by the expression for the
four-vector of the energy-momentum transferred to the unit volume per unit time:
\begin{eqnarray}
\frac{dP_\a}{dt} =
\left(  \frac {dQ}{dt}, {\vec \Im} \right) =
\frac{1}{V} \int \prod\limits_i dn_i f_i
\prod\limits_f dn_f (1-f_f)
\frac{|S_{if}|^2}{{\cal T}} q_\a , \l{dpa}
\end{eqnarray}
where $dn_i$ and $dn_f$ are the numbers of initial and final states in
an element of the phase space, $ f_i $ and $ f_f $  are the
distribution functions of initial and final particles,  $ q_\a $ is
the four-momentum transferred by neutrinos to the medium in the single reaction,
$ |S_{if}|^2/{\cal T} $ is the process S-matrix element squared per unit
time. We calculate the asymmetry of the momentum transferred to
the medium along the magnetic field direction in the processes of the
neutrino-nucleon scattering:
\begin{eqnarray}
N + \nu_i \Longrightarrow N + \nu_i  , \l{1} \\
N + \tilde\nu_i \Longrightarrow N + \tilde\nu_i ,  \l{2}
\end{eqnarray}
where $ N = n, p $ and  $ \; \nu_i = \nu_e, \nu_{\mu}, \nu_{\tau} $.
Under the condition of the remnant envelope the nucleonic gas is the
Boltzmann and nonrelativistic one. It is known that the asymmetry of the
neutrino momentum is absent in the case of $\beta$-equilibrium \cite{Kusenko}.
Thus, we discribe neutrino by the non-equilibrium local distribution function:
\begin{eqnarray}
f_{\n} = \frac{\Phi_{\n}(r,\chi)}
{\left(
\exp{(\omega / T_{\n} - \eta_{\n})} + 1
\right)} \; .
\l{fn}
\end{eqnarray}
Here  $ \chi $ is the cosine of the angle between the neutrino
momentum and the radial direction,  $ \omega $  is the (anti)neutrino
energy, $ T_{\n} $  is the (anti)neutrino spectral temperature,
$ \eta_{\n} $ is a fitting parameter( we are neglected the magnetic field
influence on the neutrino distribution function\cite{GO}).
In the calculation of $S$-matrix element of the processes
(\ref{1}) and (\ref{2}) we use the nucleon vacuum wave function with
polarizations $ S = \pm 1 $ along the magnetic field.
We also should take into account the additional interaction
energy of the nucleon magnetic moment with the magnetic field, so the
energy of the nucleon is:
%
%
$ E = m + \vec P^2 / 2m - g eB S / 2m $,
where $g$ is the nucleon magnetic
factor ($ g \!\! \simeq \!\! -1.91 $ for neutron and
$ g \!\! \simeq \!\! 2.79 $ for proton), m is the nucleon mass and
$ e>0$ is the elementary charge.

Under these assumptions we obtain the following expression for the force
density along the field:
\begin{eqnarray}
&&\Im _\| ^{(\nu)} = - \frac{ G_F^2 g } { 2 \pi }  \;
\frac{ eB }{ m_{n,p} T }  \; N_{n,p} \; N_\nu
\Bigg\{
\bigg( c_v c_a \langle \omega^3_\nu \rangle +
 c_a^2 T \langle \omega^2_\nu \rangle \bigg)
\bigg( \langle \chi^2_\nu \rangle - 1 / 3 \bigg) - \nonumber \\
&&- c_a^2  \bigg(  \langle \omega^3_\nu \rangle
- 5 T \langle \omega^2_\nu \rangle \bigg)
\bigg( 5 / 3 -  \langle \chi^2_\nu \rangle \bigg) +
2 c_a^2  J \bigg(  \langle \omega^3_\nu \rangle -
5T_\nu \langle \omega^2_\nu \rangle \bigg)
\bigg( 1 - \langle \chi^2_\nu \rangle \bigg)
\Bigg\} . \l{F} \;
\end{eqnarray}
Here $G_F$ is the Fermi constant, $c_v$ and $c_a$ are vector and axial
nucleonic current constants ($c_v \!\! = \!\! -1/2$,
$c_a \!\! \simeq \!\! -0.91 / 2 $ for neutron and
$c_v \!\! = \!\! 0.07 / 2 $,
$c_a \!\! \simeq \!\! 1.09 / 2$ for proton),
$N_{n,p}$ and $N_{\nu}$ are the local neutron (proton) and neutrino
number densities respectively,
\b
&&\langle \omega^n_\nu \rangle = N_\nu^{-1}
\int \omega^n \; f_\nu \; d^3k , \nonumber \\
&&\langle \chi^2_{\nu} \rangle  =
\left(
\int \chi^2 \;\omega \;f_{\nu} \;d^3k
\right)
\left(
\int \omega \;f_{\nu} \;d^3k
\right)^{-1}, \nonumber \\
&&J = (4 \pi)^{-1} \int \Phi_\nu(r,\chi) d \Omega \; . \nonumber
\e
In the case of an antineutrino we have to change $c_a^2 \to - c_a^2$:
\begin{eqnarray}
\Im _\| ^{(\tilde \nu)} = \Im _\| ^{(\nu)} (c_a^2 \to - c_a^2)
\end{eqnarray}
Note that the process of the neutrino scattering on protons is
supressed by the smallness of the proton number density
(under the conditions of the remnant envelope $ N_p / N_n \simeq 0.07 $).
According to Eq. (\ref{F}), the momentum asymmetry exists
if the spectral neutrino temperature differs from the medium temperature
($ T_\nu \neq T $) or the neutrino distribution is anisotropic
($ \langle \chi^2_\nu \rangle \neq 1/3 $).

In the case of the Boltzmann neutrino distribution function:
\b
f_{\n} = \Phi_{\n}(r,\chi) \exp{( - \omega / T_{\n})}  , \nonumber
\e
 the force density is simplified:
\begin{eqnarray}
\Im _\| ^{(\nu_i)} &=& - \; \frac{ 12 G_F^2 g } { \pi }  \;
\frac{ eB }{ m_{n,p} }  \; N_{n,p} \; N_\nu  \; T_\nu^2
\Bigg\{
4 c_a^2  \bigg( 2 - \langle \chi^2_\nu \rangle \bigg) + \nonumber \\
&+& 5 T_\nu / T  \bigg[
c_v c_a \bigg( \langle \chi^2_\nu \rangle - 1 / 3 \bigg) -
c_a^2  \bigg( 5 / 3 -  \langle \chi^2_\nu \rangle \bigg)  \bigg]
\Bigg\} .
\end{eqnarray}
Note, that under envelope conditions the neutrino and antineutrino parameters
for $\mu$ and $\tau$ species are approximately equal:
$ T_{\nu} = T_{\tilde \nu} , \; $
$ \langle \chi^2_{\nu} \rangle  =  \langle \chi^2_{\tilde \nu} \rangle $
\cite{J2} .
Thus, the total (neutrino and antineutrino) force density for each of
these species is simplified and can be presented in the form:
%
%
\begin{eqnarray}
\Im _\| ^{(\nu_i)} + \Im _\| ^{(\tilde \nu_i)}  =
- \frac{ G_F^2 c_v c_a g } { \pi } \;
\frac{ eB }{ m_{n,p} T } N_{n,p} \; N_\nu  \;
\langle \omega^3_\nu \rangle \;
\bigg( \langle \chi^2_\nu \rangle - 1 / 3 \bigg) ,
\end{eqnarray}
and this force density not equal to zero when
the neutrino distribution is anisotropic only. \\[2mm]

{\bf 3. Numerical estimations}\\[1mm]

%
We estimate the momentum asymmetry in the remnant envelope with the strong
toroidal magnetic field on the stage of the basic neutrino emission.
For numerical estimations we use the typical
value of the envelope density $ \rho = 5 \times 10^{11} g / cm^{3} $,
and the magnetic field strength $ B = 4.4 \times 10^{16} G $. The
(anti) neutrino parameters are taken from Ref. \citen{J2}:
\b
&&T_{\nu_e} \! \simeq \! 4  MeV, \;\;
T_{\tilde\nu_e} \! \simeq \! 5 MeV, \;\;
T_{\nu_{\mu,\tau}} \!\! \simeq \! T_{\tilde\nu_{\mu,\tau}} \!\!
\simeq \! 8 MeV,
\nonumber \\
&&N_{\nu_e} \! \simeq 5 \! \times 10^{32} cm^{-3}, \;\;
N_{\tilde\nu_e} \! \simeq \! 2.1 \times 10^{32} cm^{-3},  \;\;
N_{\nu_{\mu,\tau}} \!\! \simeq \! N_{\tilde\nu_{\mu,\tau}} \!\!
\simeq \! 1.8 \times 10^{32}  cm^{-3},
\nonumber  \\
&&\langle \chi^2_{\nu_i} \rangle \simeq
\langle \chi^2_{\tilde \nu_i} \rangle \simeq  0.4, \;\;
\eta_{\nu_i} \simeq \eta_{\tilde\nu_i} \simeq 0 .
\nonumber
\e
With these parameters we obtain the estimation of the total (summarised
over all neutrino species) force density in the neutrino-nucleon scattering:
\begin{eqnarray}
\Im_\|^{(scat)} \simeq  3.4 \times 10^{20}
\!  dynes / cm^3
 \left( \frac{B}{ 4.4 \times 10^{16} G } \right)
\! \left( \frac{ \rho }{ 5 \times 10^{11} g / cm^3 } \right).
\end{eqnarray}
Let us compare this result with the estimation of the force density
in URCA processes \cite{GO}:
\begin{eqnarray}
\Im_\|^{(urca)} \simeq  2 \times 10^{20}
\!  dynes / cm^3
 \left( \frac{B}{ 4.4 \times 10^{16} G } \right)
\! \left( \frac{ \rho }{ 5 \times 10^{11} g / cm^3 } \right).
\end{eqnarray}
We stress that these quantities are of the same sign, comparable and
sufficiently large. The total force spins up quickly the envelope along
the toroidal magnetic field direction. The estimation of the angular
acceleration:
\begin{eqnarray}
\dot \Omega \sim 10^{3} s^{-2}
\left( \frac{B}{ 4.4 \times 10^{16} G } \right)
\left( \frac{ R_c }{ 10 km } \right)
\end{eqnarray}
shows that the neutrino "spin up" effect could influence substantially on
the envelope dynamics. \\[2mm]

{\bf 4. Conclusions}\\[1mm]

In the processes of the neutrino-nucleon scattering the macroscopic
momentum is transferred to the remnant envelope along the toroidal magnetic
field direction. This momentum is large enough in the case of the strong
magnetic field and coincides in the sign with the similar momentum in the
direct URCA processes. The force which appears in the neutrino-nucleon
processes  in the toroidal magnetic field generates the torque. This
torque can quickly spin up the part of the envelope filled by the strong
magnetic field. Therefore the neutrino "spin up" effect could essentially
influence on the dynamics of the remnant envelope. \\[2mm]

{\it Acknowledgements.} The authors express the deep gratitude to the
Organizing Committee of the GMIC' 99 Conference for the possibility to
participate and for the worm hospitality. This work was supported in part
by the INTAS under Grant No. 96-0659 and by the Russian Foundation for
Basic Research under Grant No. 98-02-16694.
\\[3mm]
\indent

{\bf References\\[2mm]}

\end{document}